\documentclass[aps,prl,twocolumn,showpacs,floatfix,nofootinbib,preprintnumbers,superscriptaddress,amsmath,amssymb]{revtex4-1}
\usepackage[T1]{fontenc}
\usepackage[utf8]{inputenc}
\usepackage{graphicx}
\usepackage{epsfig}
\usepackage{bm}
\usepackage{color}
\usepackage{float}
\usepackage{dcolumn}
\usepackage{txfonts}
\usepackage{graphicx}   

\begin{document}

\title{Comment on "Direct Observation of Proton Emission in $^{11}$Be" }


\author{H.O.U. Fynbo}
\affiliation{Institut for Fysik og Astronomi, Aarhus Universitet, DK-8000 Aarhus, Denmark}
\author{Z.~Janas}
\affiliation{Faculty of Physics, University of Warsaw, 02-093 Warszawa, Poland}
\author{C.~Mazzocchi}
\affiliation{Faculty of Physics, University of Warsaw, 02-093 Warszawa, Poland}
\author{M.~Pf\"utzner}
\email{pfutzner@fuw.edu.pl}
\affiliation{Faculty of Physics, University of Warsaw, 02-093 Warszawa, Poland}
\author{J. Refsgaard}
\affiliation{Department of Astronomy and Physics, Saint Mary's University, Halifax, Nova Scotia, B3H 3C3 Canada}
\affiliation{TRIUMF, 4004 Wesbrook Mall, Vancouver BC, V6T 2A3 Canada}
\author{K. Riisager}
\affiliation{Institut for Fysik og Astronomi, Aarhus Universitet, DK-8000 Aarhus, Denmark}
\author{N.~Sokołowska}
\affiliation{Faculty of Physics, University of Warsaw, 02-093 Warszawa, Poland}

\begin{abstract}
  We argue that conclusions of [PRL 123, 082501 (2019)] are incorrect.
  The authors present the direct observation of beta-delayed proton emission in the beta
  decay of $^{11}$Be. From the determined branching
  ratio for this process and from the energy spectrum of emitted
  protons the existence of a so far unobserved narrow resonance in
  $^{11}$B was deduced. The given beta strength for the transition to this state
  is however wrong. In addition, we show that the combination
  of peak position and branching ratio is in strong disagreement with models 
  considered by the authors. Furthermore, we identify several deficiencies in 
  the analysis, and we provide possible sources of background, 
  that could explain the error.
%
\end{abstract}


\maketitle

In their recent Letter Ayyad et al. reported
the first direct observation of delayed protons emitted in the $\beta$
decay of $^{11}$Be \cite{Ayyad}. The authors claim that the
decay proceeds through a narrow resonance in $^{11}$B at an excitation
energy of 11.425(2) MeV with a branching ratio of
$1.3(3) \times 10^{-5}$. From the position of the resonance and the
measured branching ratio Ayyad et al. calculate the log($ft$) value to
be 4.8(4). This value is wrong. A simple estimate based on the resonance position
and experimental branching ratio gives a log($ft$) value of 2.9 for a
sharp resonance. Inserting a realistic width will not change this
value much, which means that the beta strength $\langle GT \rangle^2$
would be around 4.8, in significant disagreement with the model
presented in Ref.\ \cite{Ayyad}. Such a large value cannot be
explained in any single-nucleon model where $\langle GT \rangle^2$ at
most can be 3. The combination of the
resonance position and branching ratio reported by Ayyad et al. is
in contradiction with any model presented so far.

The direct observation of protons emitted in the $\beta$
decay of $^{11}$Be is extremely challenging because of the small
energy window available of about 280 keV and the very low branching
ratio \cite{Borge}. Riisager et al. \cite{Riisager} reported the
indirect observation of this decay channel with a branching ratio of
$(8.3 \pm 0.9)\times 10^{-6}$ through identification with the AMS
technique of atoms of $^{10}$Be in collected samples of $^{11}$Be.
Importantly, $^{11}$Be may also decay by $\beta$-delayed $\alpha$ emission
($\beta \alpha$) \cite{Refsgaard} with the much larger total branching
of 3.30(10)\%. In the experiment of Ayyad et al. the selected ions
of $^{11}$Be were stopped inside a gaseous time projection chamber.
About 90\% of them drifted to the
cathode before decaying - in such case only one heavy decay product
could enter the active volume and be recorded.

\begin{figure}
  \resizebox{0.38\textwidth}{!}{%
    \includegraphics{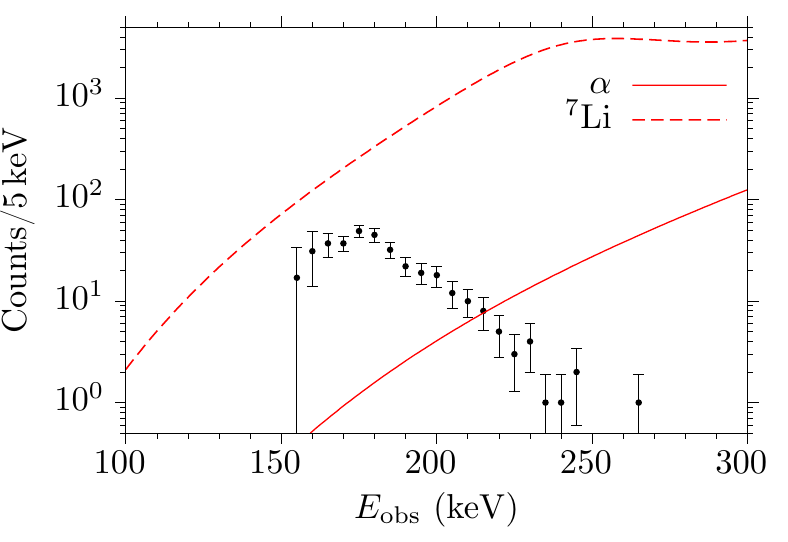}
  }
  \vspace*{-5mm}
\caption{The $\beta$-delayed proton spectrum from $^{11}$Be reported
  in Ref.\ \protect\cite{Ayyad} is shown along with extrapolated
  intensities of $\alpha$ particles and $^7$Li recoils from the
  $\beta\alpha$ branch as determined in Ref.\ \protect\cite{Refsgaard}.}
\label{fig:spectra}
\end{figure}

The most likely cause of error is a contribution to the spectrum
from particles other than protons.  A key ingredient of the analysis
was the selection of candidate proton events from a large background
coming from the $\beta \alpha$ channel.  We show in
Fig. \ref{fig:spectra} the expected spectra of recoiling $^{7}$Li ions
and $\alpha$ particles at low energy based on fits obtained in a
recent experiment \cite{Refsgaard}. In their analysis Ayyad et
al. entirely neglected low-energy $\alpha$ particles as a possible
source of background.  However, the extrapolation shown in
Fig. \ref{fig:spectra} indicates that about 150 $\alpha$ particles
could be expected below 250 keV, which amounts to 40\% of the reported
branching ratio.
In Ref.\ \cite{Ayyad} the $^{7}$Li ions were taken into account as a
source of background, but the quality of the proton-$^7$Li
discrimination was not documented; Fig. 2 of Ref.\ \cite{Ayyad} shows
only two example events of unknown energy. Using the absolute values
of $\chi ^2$, as in their Fig. 3, is not appropriate as they depend on
the number of data points (samples) fitted, and thus on the particle
energy and the emission angle. For event classification the normalized
$\chi ^2$ per number of degrees of freedom should have been used
instead. Hence, a fraction of $^7$Li ions could also be a source of background
leading to the unphysical peak position and branching ratio reported by Ayyad et
al.

An additional issue with the analysis of Ayyad et al. concerns the
extraction of the width of the resonance.  The width of the proton
peak shown in Fig. 3 in Ref.\ \cite{Ayyad} is 42 keV (FWHM), which is
inconsistent with the quoted energy resolution of 15 keV and the width
of the fitted Breit-Wigner distribution ($\Gamma = 15\,$ keV).


%

\end{document}